# Autofocusing technologies for whole slide imaging and automated microscopy

*Zichao Bian[1,3], Chengfei Guo[1,3], Shaowei Jiang[1,*], Jiakai Zhu[1], Ruihai Wang[1], Pengming Song[2], Zibang Zhang[1], Kazunori Hoshino[1], and Guoan Zheng[1,*]*

[1]University of Connecticut, Department of Biomedical Engineering, Storrs, CT, 06269, USA
[2]University of Connecticut, Department of Electrical and Computer Engineering, Storrs, CT, 06269, USA
[3]These authors contributed equally to this work
*E-mail: shaowei.jiang@uconn.edu (S. J.) or guoan.zheng@uconn.edu (G. Z.)

**Abstract:** Whole slide imaging (WSI) has moved digital pathology closer to diagnostic practice in recent years. Due to the inherent tissue topography variability, accurate autofocusing remains a critical challenge for WSI and automated microscopy systems. The traditional focus map surveying method is limited in its ability to acquire a high degree of focus points while still maintaining high throughput. Real-time approaches decouple image acquisition from focusing, thus allowing for rapid scanning while maintaining continuous accurate focus. This work reviews the traditional focus map approach and discusses the choice of focus measure for focal plane determination. It also discusses various real-time autofocusing approaches including reflective-based triangulation, confocal pinhole detection, low-coherence interferometry, tilted sensor approach, independent dual sensor scanning, beam splitter array, phase detection, dual-LED illumination, and deep-learning approaches. The technical concepts, merits, and limitations of these methods are explained and compared to those of a traditional WSI system. This review may provide new insights for the development of high-throughput automated microscopy imaging systems that can be made broadly available and utilizable without loss of capacity.

**Keywords**: whole slide imaging, digital pathology, focus quality, focus map, deep learning.

# 1. Introduction

The process of analyzing pathology slides using an optical microscope has remained relatively unchanged until recently. In a regular process, pathologists move the microscope stage to different positions to identify areas of interest, which can be further analyzed by switching to a higher magnification objective lens. The focusing of the slide is manually performed using the focus knob of the microscope platform. Although this traditional slide reviewing process remains the gold standard in diagnosing a large number of diseases including almost all types of cancers, it is highly subjective on the other hand: different pathologists may arrive at different conclusions and the same person may also give different conclusions at different time points. In terms of workflow efficiency, this process is labor-intensive and can be easily disrupted when a pathologist bumps a slide to a high magnification objective lens[1]. Similarly, it can be disrupted when the pathologist switches to a different objective lens and performs manual focusing of the slide. After the reviewing process, the slides must be kept accessible, clean and protected, creating additional storage and labor demands[1,2].

Since the current slide reviewing process is based on subjective opinions of pathologists, there is a need for quantitative and streamlined assessment of histology slides. Quantitative characterization of pathology imagery is not only important for reducing inter- and intra-observer variations in diagnosis but also to better understand the biological mechanisms of the disease process[3]. Recent clinical guidelines have begun to require quantitative evaluations as part of the effort towards better patient risk stratification[4]. For example, breast cancer staging requires the counting of mitotic cells.

A whole slide imaging (WSI) system is designed to replace the traditional microscope for quantitative and streamlined slide reviewing. It was first developed based on a robotic microscope platform in the 1990s[5]. The essential components of a WSI system include the following: 1) a microscope with objective lenses, 2) robotics to move slides, 3) one or more image sensors for image acquisition and autofocusing, and 4) software for management. In the acquisition process, a typical WSI system captures hundreds of high-resolution images that are subsequently aligned and stitched together to create a complete and seamless representation of the original whole tissue section[6]. The stitched whole slide image can provide a digital equivalent of the original glass slide on the microscope. The pathologists can then view, navigate, change magnification, and annotate the virtual slide with speed and ease. Digital pathology using WSI is now advancing into clinical workflow for better and faster predication, diagnosis, and prognosis of cancers and other diseases[1]. A major milestone was accomplished in 2017 when the U.S. Food and Drug Administration approved the first WSI scanner for primary diagnostic use in the U.S.[7,8]. The new generation of pathologists trained on digital pathology promises further growth of the field in the coming decades.

Another driving force for the development of digital pathology is the recent advancement of artificial intelligence (AI) in medical diagnosis[9-13]. In particular, deep-learning approaches have been demonstrated for automated analysis of microscopic pathology images with performance comparable to that by human experts[14-18]. An augmented reality microscope has also recently been developed to provide real-time integration of AI in the slide inspection process[15]. In this augmented reality microscope platform, two modules are attached to a regular upright microscope. The first module is a digital camera that captures high-resolution images of the same field of view as one observes through the eyepiece ports. The second module is a microdisplay that projects digital information into the eyepiece ports. In a typical implementation, the captured image from the camera will be processed by a deep learning algorithm to produce a heatmap that predicts tumor probability. The outline of the predicted tumor regions will then be projected to the eyepiece ports via the microdisplay. As such, the pathologists can observe the original specimen overlaid with the AI-assisted information through the eyepiece ports.

A fundamental challenge with WSI, automated microscopy, and augmented reality microscopy has been the ability to acquire high-quality, in-focus images at high speed. For a high numerical aperture (NA) objective lens, the depth of field is on the orders of 1 µm. The small depth of field poses a difficulty to track the axial topography variations that inherently exist in solid tissue samples[6]. If the specimen is not placed within the depth of field of the objective lens, the image quality of the acquisition will be degraded, leading to rescanning and workflow delays. Several studies have implicated poor focus as the main culprit for poor image quality in WSI[19-21]. For augmented reality microscopy, defocus blur can occur to the captured images due to the optical path length difference between the eyepiece ports and the camera port. This optical path length difference varies for different objective lenses. As a result, it is challenging to maintain the in-focus position for the camera when the pathologist keeps switching to different objective lenses in the slide reviewing process. Furthermore, some pathologists may have certain vision conditions such as myopia. Instead of adjusting the diopter on the eyepieces, they may prefer to adjust the focus knob to bring the sample into focus for their eye observation. The captured image through the camera port, on the other hand, will be out-of-focus due to the introduced optical path length difference. To address these challenges in augmented reality microscopy, a real-time autofocusing module is needed to acquire high-quality, in-focus images at high speed.

Here we review and discuss different autofocusing techniques for WSI and automated microscopy in general. A list of commercially-available WSI scanners and automated microscopy systems are provided in Table 1. The employed autofocusing techniques are listed in the last column and they can be categorized into three groups: 1) pre-scan focus map approach, 2) real-time reflective autofocusing, and 3) real-time image-based autofocusing. In the following, we will first review the traditional pre-scan focus map approach in Section 2. We will discuss the choice of different focus measures for determining the best focal position. In Section 3, we will review the reflective autofocusing approaches, including intensity detection via confocal pinhole, triangulation with oblique illumination, and low-coherence interferometry. In Section 4, we will review and discuss various real-time image-based autofocusing approaches, including tilted sensor, independent dual sensor scanning, beam splitter array, phase detection, dual-LED illumination, and deep-learning approaches. The technical concepts, merits, and limitations of these methods are explained and compared to those of a traditional focus map approach. In Section 5, we will summarize our discussion and provide perspectives for future development. This review may provide new insights for the development of high-throughput automated microscopy systems that can be made broadly available and utilizable without loss of capacity.

| Vendor | Model | Imaging mode | Slide capacity | Scanning speed (15 mm x 15 mm region) | Sensor type | Autofocusing method |
|---|---|---|---|---|---|---|
| Zeiss | Axio Scan.Z1 | Brightfield, Fluorescence | 12 or 100 slides | 20× 240 sec/slide | 3 CCD sensor, sCMOS sensor | Focus map |
| Olympus | VS200 | Brightfield, Darkfield, Phase contrast, Polarization, Fluorescence | 210 slides | 20×: 80 sec/slide | Area sensor | Focus map |
| Hamamatsu | NanoZoomer S360 | Brightfield | 360 slides | 20×: ~30 sec/slide 40×: ~30 sec/slide | TDI sensor | Focus map |
| Huron | TissueScope LE120 | Brightfield | 120 slides | 20×: <60 sec/slide | Area sensor | Focus map |
| Ventana | iScan HT | Brightfield | 360 slides | 20×: <45 sec/slide 40×: <72 sec/slide | Information not available | Focus map |
| Leica | Aperio AT2 DX | Brightfield | 6 or 400 slides | 20×: <72 sec/slide | TDI sensor | Focus map |
| Leica | Aperio GT 450 | Brightfield | 450 slides | 40×: 32 sec/slide | TDI sensor | Tilted sensor |
| 3DHistech | Pannoramic 1000 | Brightfield | 1000 slides | 20×:<60 sec/slide 40×:<60 sec/slide | Area sensor | Focus map |
| 3DHistech | Pannoramic 250 Flash III | Brightfield, Fluorescence | 250 slides | 20×: 35 sec/slide 40×: 95 sec/slide | 3 CCD sensor, sCMOS sensor | Focus map |
| Philips | Ultra fast scanner | Brightfield | 300 slides | 40×: 60 sec/slide | TDI sensor | Tilted sensor |
| Nikon | Eclipse Ti2 (Perfect Focus) | Brightfield, Phase contrast, Fluorescence | 1 slide | Not available | Area sensor | Triangulation with oblique illumination |
| Olympus | IXplore (TruFocus) | Brightfield, Phase contrast, Fluorescence | 1 slide | Not available | Area sensor | Triangulation with oblique illumination |
| Thorlabs | EV103 | Brightfield, Fluorescence | 4 slides | 20×: <70 sec/slide 40×:<200sec/slide | TDI sensor | Low-coherence interferometry |
| Omnyx (now Inspirata) | VL120 | Brightfield | 120 slides | 40×: 80 sec/slide 60×: 200 sec/slide | Area sensor | Independent dual sensor scanning |

**Table 1.** A list of commercially-available WSI scanners and automated microscopy systems. Note: every attempt was made to include accurate data in this table at the time of writing this article. The autofocusing methods were best estimated based on the product instruction manuals and related patents.

## 2. Focus map surveying

Focus map surveying is the most adopted autofocusing method in commercially available WSI systems. Manufactories are in favor of using this approach because of two main reasons: 1) it requires no additional optical hardware and is robust for different types of samples, and 2) no or less intellectual property issue. Here we will first discuss the choice of focus measure in Section 2.1. We will then discuss focus map generation and focus quality control in Section 2.2.

## 2.1. Z-stack acquisition and focus-measure calculation

The principle of this method is shown in Figure 1, where the camera is used to acquire z-stack images of the specimen when the sample or the objective lens is axially scanned to different positions. From the resulting z-stack, a certain figure of merit of each image, such as image contrast, entropy, spatial frequency content, is extracted for measuring the quality of focus. It is also common to acquiring images while calculating the figure of merit, and choosing the image corresponding to the peak (or valley) of the figure of merit, or by performing a search to optimize the figure of merit. By repeating this searching process for different tiles of the microscope slide, the well-focused digital whole slide image can be obtained.

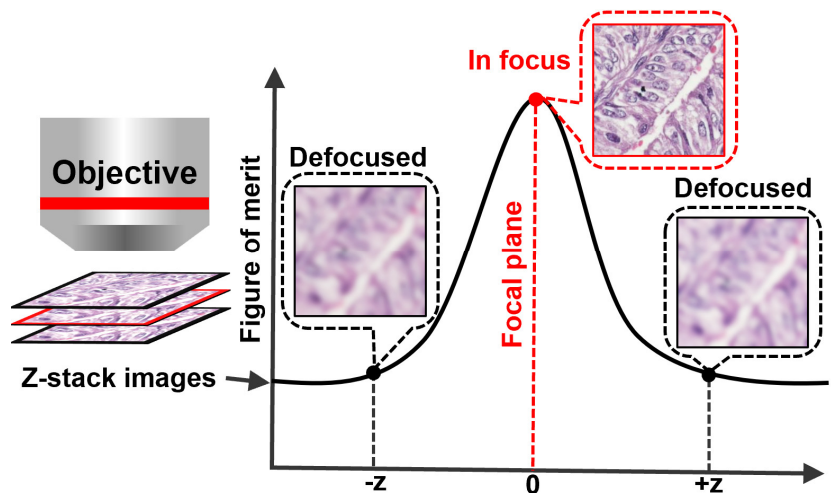

**Figure 1.** The traditional axial scanning procedure for autofocusing. For a selected region of interest, a z-stack is acquired and used to determine the focus position using a certain figure of merit.

An important aspect of this approach is to choose a proper figure of merit to measure the quality of focus. When the specimen is in focus, the captured image should demonstrate large image contrast, a large range of intensity values, and sharp edges. Quantitatively, a good figure of merit should be acutely sensitive to focus, monotonically decreasing and symmetric about the peak, and contains no prominent local maxima outside of the peak, as shown in Figure 1. Accuracy is clearly of utmost importance. In the case of WSI and automated microscopy, minimizing the computation time is also critical.

Several previous studies have evaluated and compared a list of common focus measures[22-27]. Table 2 lists a dozen common focus measures that are intuitive and computationally simple. In general, they can be categorized into 4 groups[23]: (1) derivative-based measures such as Brenner gradient, Tenenbaum gradient, energy Laplace, Gaussian derivative, sum of wavelet coefficients, ratio of wavelet coefficients, power-weighted average, and power log-log slope, (2) statistical-based measures such as image contrast, normalized variance, auto-correlation, and standard deviation-based correlation, (3) histogram-based measures such as histogram range, histogram entropy, and weight histogram sum, and (4) intuitive-based measures such as thresholded content.

| Focus measure | Equation | Comments |
|---|---|---|
| Brenner gradient[28] | $F_{Brenner} = \sum_x \sum_y (I(x+2,y) - I(x,y))^2$, where $I(x,y)$ is the captured 2D intensity image. | High autofocusing accuracy for different samples[23,25]. |
| Tenenbaum gradient[29] | $F_{Tenengrad} = \sum_x \sum_y (S_x(x,y)^2 + S_y(x,y)^2)$, where $S_x(x,y)$ and $S_y(x,y)$ are the resultant images by convoluting $I(x,y)$ with the kernels [-1 0 1; -2 0 2; -1 0 1] and [1 2 1;0 0 0;-1 2 -1], respectively. | Well performed for subsampled images and robust to random noises[23,30]. |
| Energy Laplace[31] | $F_{energy\_Laplace} = \sum_x \sum_y [I(x-1,y) + I(x+1,y) + I(x,y-1) + I(x,y+1) + 4I(x-1,y)]^2$ | Well performed for tuberculosis detection[30,32]. |
| Gaussian derivative[33] | $F_{Gaussian} = \frac{1}{X \cdot Y} \sum_x \sum_y [I(x,y) * G_x(x,y,\sigma)]^2 + [I(x,y) * G_y(x,y,\sigma)]^2$, where $G_x$ and $G_y$ are the first-order Gaussian derivatives in x- and y-direction at scale $\sigma$. | Robust against noises with a proper selection of parameter $\sigma$33. |
| Sum of wavelet coefficients[34,35] | $F_{sum\_wavelet} = \sum_\omega |W_{HL}(x,y)| + |W_{LH}(x,y)| + |W_{HH}(x,y)|$, where $\omega$ is the corresponding window in the DWT sub-regions. $W_{HL}, W_{LH}$ and $W_{HH}$ are the level-1 two-dimension DWT sub-regions. | A common derivative-based focus measure[34,35]. |
| Ratio of wavelet coefficients[36] | $F_{ratio\_wavelet} = M_H^2/M_L^2$, $M_H^2 = \sum_K \sum_{xy} W_{HLn}(x,y)^2 + W_{LHn}(x,y)^2 + W_{HHn}(x,y)^2$, $M_L^2 = \sum_{xy} W_{LLk}(x,y)^2$, where $W_{LLk}$ is the Kth level DWT low-frequency sub-region. $W_{HLn}, W_{LHn}$ and $W_{HHn}$ are the level-n two-dimension DWT sub-regions. | Well performed for common microscopic images[36]. |
| Power-weighted average[37,38] | $F_{index}(z) = \sum_x \sum_y [f(x,y) * I_z(x,y)]^2 / [\sum_x \sum_y I_z(x,y)]^2$ and $F_{power\_weight} = \sum_z z F_z(z)^m / \sum_z F_z(z)^m$, where $f(x,y)$ is high-pass or band-pass filter, $*$ stands for the convolution operator, $I_z(x,y)$ is the grey level intensity of pixel (x, y) at z position. $m$ is an integer chosen by the user for different applications. | Well performed for phase-contrast autofocusing[37-41]. |
| Power log-log slope[42] | $F_{PLLS}$ is the log-log slope of the one-dimensional power spectral density $F_{PSD}$ of image $I$, where $F_{PSD} = \log(abs(FT(I))^2)$ and $FT$ denotes as Fourier transform. | Well performed for focus quality control in high-content screening[42,43]. |
| Image contrast[25] | $F_{contrast} = (I_{max} - I_{min})/(I_{max} + I_{min})$, where $I_{max}$ and $I_{min}$ are the maximum and minimum grey level intensity, respectively. | A common statistical-based focus measure[25]. |
| Normalized variance[22] | $F_{normed\_variance} = 1/(X \cdot Y \cdot \mu) \sum_x \sum_y (I(x,y) - \mu)^2$, where $\mu$ is the mean gray level of the image. | Well performed for blood smear and pap smear autofocusing[23,24,26]. |
| Auto-correlation[44,45] | $F_{autocorr} = \sum_x \sum_y I(x,y) \cdot I(x+1,y) - \sum_x \sum_y I(x,y) \cdot I(x+2,y)$ | Well performed for fluorescence microscopy[23,46]. |
| Standard deviation-based correlation[44,45] | $F_{corr\_stddev} = \sum_x \sum_y I(x,y) \cdot I(x+1,y) - X \cdot Y \cdot \mu^2$ | Robust to noises[23]. |
| Histogram range[47] | $F_{range} = \max_I(h(I) > 0) - \min_I(h(I) > 0)$, where $h(I)$ is image histograms (i.e., the number of pixels with intensity $I$ in an image). | Performance depends on samples and imaging methods[23,47]. |
| Histogram entropy[47] | $F_{entropy} = -\sum_I p_I \cdot \log_2(p_I)$, where $p_I = h(I)/(X \cdot Y)$ is the probability of a pixel with intensity $I$. | Well performed for sinusoidal and binary images[47]. |
| Weight histogram sum[26,48] | $F_{WHS} = \sum_I [\sqrt[5]{h(I)} \cdot I(x,y)^5 \cdot 10^{-15}]$, where the fifth root and fifth potency are empirical results. | Well performed for fluorescence bacterial samples[26,48]. |
| Thresholded content[22,49] | $F_{th\_cont} = \sum_x \sum_y I(x,y)$, where $I(x,y) \geq \theta$. $\theta$ is the threshold | Fast computation; a good choice for the coarse searching[26]. |

**Table 2.** Common figure of merits for measuring the focus quality.

With a chosen focus measure for certain applications, the next step is to estimate the focus position using the calculated focus measure from the acquired images. A fitted function can be used to find the peak (or valley) from the figure of merit data points, obviating the need to acquire images near the focus. The choice of curve fitting model directly affects the number of images needed. Typical fitting models include polynomial[50], Lorentzian[25], and Gaussian models[51,52]. A polynomial fit may closely approximate the figure of merit data points that are close to the focal plane. An $n^{th}$-order function, however, requires a minimum of $n+1$ images to be acquired, thus drastically increasing image acquisition time when a higher-order fitting curve is employed. It may also fail if the focus plane is substantially outside of the depth of field. Yazdanfar *et al.* have demonstrated a Lorentzian function for fitting the Brenner gradient focus measure[25]. Using this empirical model, only 3 images are needed to determine the focal plane. Similarly, Gaussian fitting model with 3 unknown parameters has been demonstrated for fluorescence microscopy with an electrically tunable lens[52]. The choice of fitting model is an important topic for each of the chosen focus measure and the related microscopy applications. Further research in this direction is highly desired.

The focus measures listed in Table 2 are mainly designed for incoherent microscopy with intensity-only measurements. Another important property of light wave is phase, which characterizes the optical delay accrued during propagation. Light detectors such as image sensors and photographic plates can only measure intensity variation of the light waves. Phase information is lost during the recording process. Consequently, phase measurement often involves additional experimental complexity, typically by requiring light interference with a known field[53,54], or via a phase retrieval process where the complex amplitude is recovered from intensity measurements[55].

Coherent microscopy uses both intensity and phase as the focus measure. The autofocusing process can be performed after the data has been acquired. As one example, Fourier ptychography is a coherent microscopy technique that has been demonstrated for WSI[56]. Unlike in conventional microscopy where resolution and imaging field of view need to be traded off against each other, Fourier ptychography can achieve both high resolution and wide field of view via a low-NA objective lens and angle-varied illumination. Regular WSI platform stitches the captured intensity images in the spatial domain to expand the field of view. Fourier ptychography, on the other hand, stitches the information in the Fourier domain to expand the spatial frequency bandwidth. Autofocusing of Fourier ptychography is performed in the ptychographic phase retrieval process[57,58], where the defocus pupil aberration can be jointly recovered with the complex object[59-61]. At the end of the reconstruction, the synthesized information in the Fourier domain generates a high-resolution, complex-valued object image that retains the original large field of view set by the low-NA objective lens. Similar coherent imaging procedures can also be performed at the detection path via aperture or diffuser modulation[62-67]. In this case, the recovered complex wavefront can be digitally propagated to any plane along the optical axis after reconstruction. A focus measure with both intensity and phase can be used to determine the best focal plane of the object[68-78]. A detailed discussion on coherent microscopy and the related focus measures are beyond the scope of this review article. In the following, we focus our discussions on regular incoherent microscopy.

## 2.2. Focus map, skipping tiles, and focus quality control

By repeating the z-stack autofocusing process for every tile, it is straightforward to generate a high-resolution, well-focused whole slide image of the specimen. However, as indicated above, the autofocusing process can take a significant amount of time to acquire z-stacks at multiple positions. Assuming a rate of 20 frames per second to acquire images, surveying focus at 5 different focal positions would take 0.25 seconds per tile. As a result, an image with 500 tiles can take as much as 150 seconds to acquire, not including the deceleration, acceleration, settling time for moving the slide to different lateral and axial positions. Therefore, it is not a feasible solution to perform autofocusing on every tile using the traditional image-based focus measure approach. To address the time burden, many WSI systems create a focus map prior to scanning, or survey focus points every *n* tiles or lines, in effect skipping areas to save time[6]. The number and the locations of the focus points are often made user selectable.

Figure 2(a) shows the procedures of the focus map surveying approach. The system will first select focus points based on the sample's feature and distribute them evenly over the entire slide. Each focus point is triangulated to create a focus map of the tissue surface, in effect filling in the blanks. Delaunay triangulation is a typical method for generating the focus map[6]. As shown in Figure 2(b), line scanners typically achieve better autofocusing performance than traditional 2D area sensors because linear sensors can change focus at a shorter interval.

Regular 1D and 2D image sensors need to have high illumination intensity to quickly register light levels before the sample motion causes smearing of the image. Time delay integration (TDI) sensor overcomes this illumination limitation by having multiple rows of elements that each shift their partial measurements to the adjacent row synchronously with the motion of the image across the array of elements[41]. TDI sensors are often the choice of low-light applications such as fluorescence microscopy with low photon budgets. The disadvantage of TDI sensor is the requirement of precisely synchronized sample scanning for generating an image. Rescan of the sample is needed for imaging multiple depths or fluorescence channels. Precise co-localization of different depths or different fluorescence colors can be a challenge for post-acquisition processing. The use of TDI sensors also lacks the imaging flexibility for research microscopy in general.

An alternative approach to generating the focus map is to perform autofocusing in every *n* tiles, termed 'skipping tiles' in Figure 2(b). In this case, it assumes the focused tile shares the same focus position with its adjacent tiles. The focusing performance is, however, worse than the focus map approach as it may contain more out-of-focus regions as shown in Figure 2(b). The skipping tiles approach, on the other hand, does not need to travel back to a certain axial position with sub-micron accuracy. The requirement of motion repeatability is not as stringent as that in the focus map approach. Nevertheless, more focus points can increase the accuracy of the overall focusing performance for both approaches, at the expense of additional time for autofocusing.

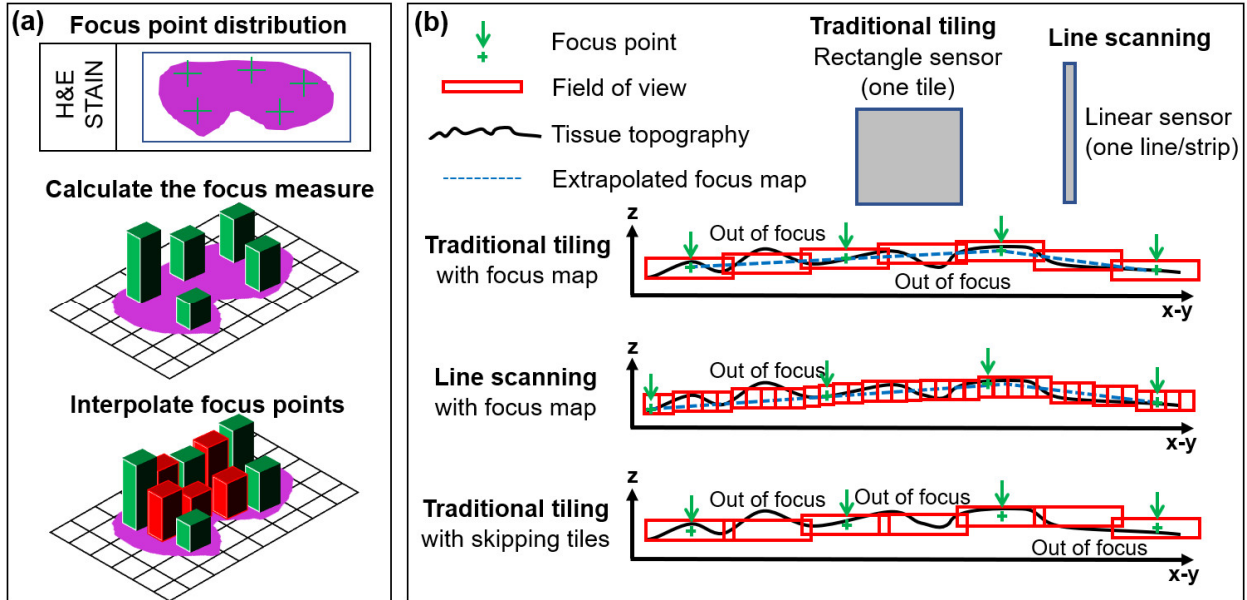

**Figure 2.** (a) Focus map generation procedures. The green bars represent the calculated figure of merits at different focus points. The red bars represent the interpolated focus points. (b) Comparison between the focus map approach and the skipping tiles approach. The green crosshairs represent the focus points used to calculate the focus map. The blue dashed lines represent the focus positions interpolated between the selected focus points. Red boxes represent the focal plane for each field of view using a 2D image sensor or a 1D linear sensor. Each red box can be adjusted in the z-axis during a scan. Modified from Ref. [6].

After the high-resolution specimen images are acquired, it is often necessary to review the images for focus quality control and determine whether certain regions need to be re-scanned. Similarly, in high-content screening for drug discovery and genome analysis, it is important to identify out-of-focus images for obtaining a clean, unbiased image dataset. Complicating this task is the fact that one only has a single-z-depth image instead of a z-stack for analysis. An absolute measure of image focus on a single image in isolation, without other user-specified parameters, is needed in this case. In the past years, various approaches have been demonstrated for no-reference focus quality assessment, including gradient map[79-81], contrast map[82-84], phase coherency[85,86], cumulative probability of blur detection[87,88], visual system's equalization of spatial frequency[89], among others. Jimenez *et al.* have tested several quality assessment metrics on a database of pathology slides and reported that the cumulative probability of blur detection is most effective among the 6 tested metrics[86]. Another emerging direction for focus quality control is to convert the image assessment process into a classification task using a neural network[21,90-94]. For example, Senaras *et al.* reported a 'DeepFocus' network to identify out-of-focus regions in histopathological images[92]. Discussions of deep-learning approaches will be given in Section 4.6.

## 3. Reflective-based autofocusing

Reflective-based autofocusing aims to detect the axial location of a reference plane, which is usually the interface between glass and liquid where the cells residue or the air-glass interface at the bottom of the cell culture vessels. During experiments, the focus drift correction system will repetitively find the axial location of the reference plane and maintain a constant distance between the objective lens and the reference plane through a motorized axial driver. In Section 3.1, we will discuss a confocal pinhole approach to locate the interfaces. In Section 3.2, we will discuss how to use the reflective light displacement to locate the reference plane in real-time. In Section 3.3, we will discuss a low-coherence interferometry approach to locate the sample switched by two interfaces in real-time.

## 3.1. Confocal pinhole detection

Liron *et al.* reported a laser reflective autofocusing approach using confocal pinhole detection in 2006[95]. The optical setup is shown in Figure 3, where a laser beam is expanded and focused onto the substrate of the sample (highlighted in red). The reflective light from the substrate passes through a confocal pinhole and reaches the photodetector (highlighted in yellow). The fraction of laser intensity reflected at an interface is roughly proportional to the square of the refractive index difference. For biological specimens located in water (or aqueous buffers) above a glass / plastic plate, the reflection from the glass-air interface is about 4% of the incident beam and the reflection from the glass-water interface is only 0.4%. The inset of Figure 3 shows a measured intensity curve by axially scanning the objective lens to different positions. The first strong peak corresponds to the air-glass interface and the second weaker peak corresponds to the sample-glass interface. Solid and dashed lines are results for 100-µm and 200-µm pinhole. Increasing the confocal pinhole size can broaden the width of the peaks as indicated by the dashed line in Figure 3. This adjustment can reduce some unwanted interference speckles and facilitate the data analysis process. A two-stage operation was employed to perform the autofocusing process. The first stage, termed 'long peak detection search', is to locate the strong peak via high-speed axial scanning of the objective lens. With the location of the first strong peak, the position of the second peak can be estimated by adding the thickness of the glass substrate. The second stage, termed 'local peak search', performs precise peak search over a relatively short range.

While this confocal detection approach can perform precise autofocusing, its main drawback is the requirement of axial scanning to get the trace curve shown in Figure 3. Another drawback is the orders of magnitude difference in strength for the two peaks. The weaker peak can easily be overwhelmed by the first strong peak, especially for lower magnification objective lenses. In Section 3.2, we will discuss a strategy to address the first drawback, i.e., to locate the first peak position without performing axial scanning. In Section 3.3, we will discuss another strategy to address both drawbacks, i.e., to reduce the signal strength from the first peak and to locate both peaks without axial scanning.

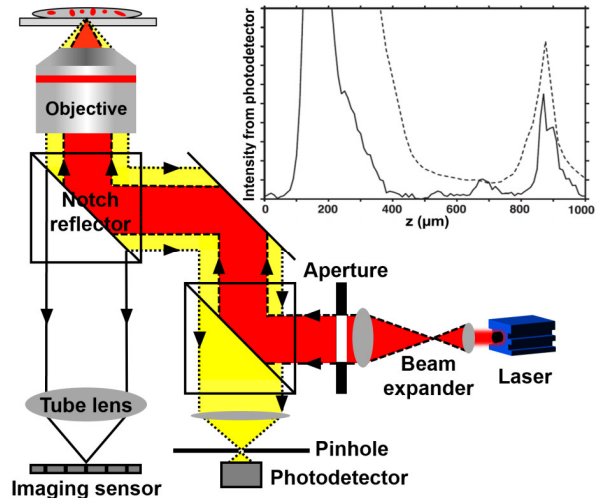

**Figure 3.** An autofocusing system using confocal pinhole detection. Laser light is expanded and focused on the substrate of the sample. The reflective light, highlighted in yellow, is passed through a confocal pinhole and detected by the photodetector. Inset in the top right shows the measured intensity signals by axially scanning the objective lens to different positions. The first strong peak corresponds to the air-glass interface and the second weaker peak corresponds to the sample-glass interface. Solid and dashed lines are results for 100 μm and 200 μm pinhole. Modified from Ref. [95].

## 3.2. Triangulation with oblique illumination

To locate the axial position of an interface without axial scanning, one can illuminate the sample with a tilted incident angle and measure the lateral displacement of the reflected beam. The triangulation concept for microscopy autofocusing can be dated back to the patent by Reinheimer in 1973[96]. In this patent, Reinheimer proposed to restrict a shaped illumination beam to occupy only half of the pupil aperture cross-section. As such, the beam reflected from a surface will have different lateral displacements when the sample surface is placed at different axial positions. The reflected light from the sample surface is detected by two photoelectric transducers for differential measurement. The differential signal detected by these two transducers is used to drive the focus knob. For example, if the sample surface is placed at the in-focus position, the reflected light will be directed to the boundary of the two transducers. The resulting differential signal is 0 and no adjustment is needed. If the sample surface is positioned above the in-focus plane, the reflected light will shift to one of the transducers. The differential signal is then used to drive down the sample stage. Similarly, if the sample surface is positioned below the in-focus plane, the differential signal from the two transducers drives up the sample stage. There are some further refinements and developments of this original patent in the 1980s and 1990s[97-104]. These developments are, in general, about how to better detect the beam size and the positional shift to infer the defocus distance. Similar schemes have also been reported in more recent literatures[105-110].

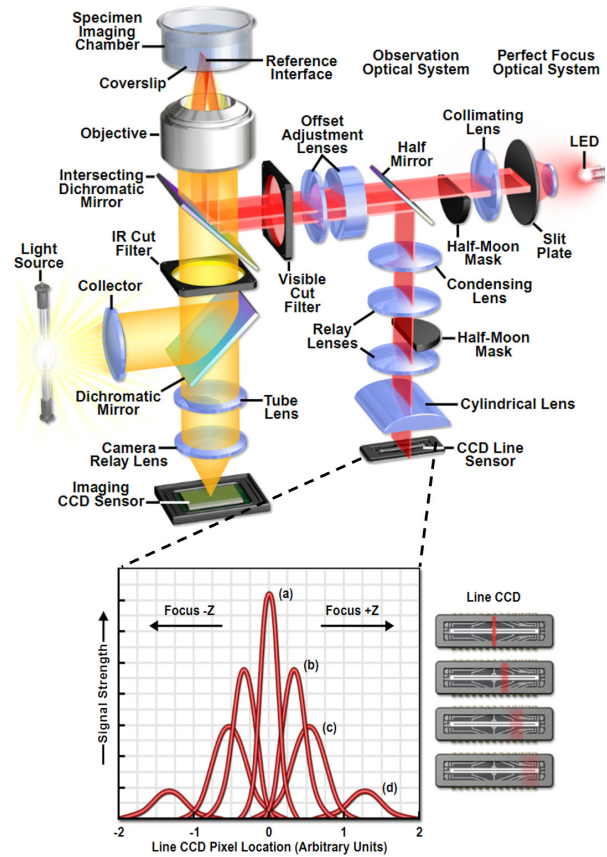

**Figure 4.** The Nikon Perfect Focus System. Light from an infrared LED is shaped by a line aperture and a half-moon mask for illuminating the sample substrate at a tilted angle. The reflected light is detected by a linear CCD. Inset shows the detected line traces when the sample substrate is scanned to different defocused positions. Two offset adjustment lenses are used to maintain the focus at the desired positional offset from the coverslip surface. Modified from Ref. [111].

Figure 4 shows the adoption of the triangulation idea in a modern microscope system, marketed as Nikon Perfect Focus System (PFS)[111]. This system maintains focus by detecting and tracking the position of the coverslip surface in real-time. It employs a near-infrared 870-nm LED as the light source and a linear CCD sensor as the detector (other detectors such as four-quant photodiode and area sensor can also be used here). Predefined by the user is an offset between the reference plane and the axial location of the desired focused image. Different from the original patent by Reinheimer, the PFS system introduces two offset adjustment lenses in Figure 4 to maintain the focus at the desired positional offset from the coverslip surface. When the user changes the offset distance, the distance of the two offset adjustment lenses changes, resulting in a shift of the line position detected by linear CCD (inset of Figure 4). The positional shift generates a signal to move the objective lens along the axial direction until the line position is centered at the linear CCD again.

The PFS system is mainly designed to image living cells housed in imaging chambers equipped with a standard coverslip. For fixed specimens like tissue slides that are mounted in a high refractive index medium (which closely matches that of the coverslip), the refraction index difference may not generate sufficient signal to detect the interface surface. Likewise, tissue slides with strong absorption profiles often scatter a considerable amount of light, leading to excessive hunting or errors in the focus drift correction system. Plastic tissue culture dishes are also not recommended, as the boundary surface may not be detectable due to insufficient offset[111].

### 3.3. Low-coherence interferometry with oblique illumination

The idea of using optical coherence tomography (OCT) for autofocusing was proposed in a patent by Wei and Hellmuth in 1996[112]. The general concept is to locate the sample position using the axial depth reflectivity profile called A-scan, which contains scattering information of sample structures along the axial direction. In the original patent, an on-axis configuration is used to perform autofocusing of an ophthalmologic surgical microscope. However, it is not suitable for high-resolution imaging of tissue slides covered by glass. The main difficulty is the overlap between the large signal reflected by glass surfaces and the weak signal reflected from the sample. Locating the sample position with submicron accuracy is challenging given the large signals reflected from the glass surfaces.

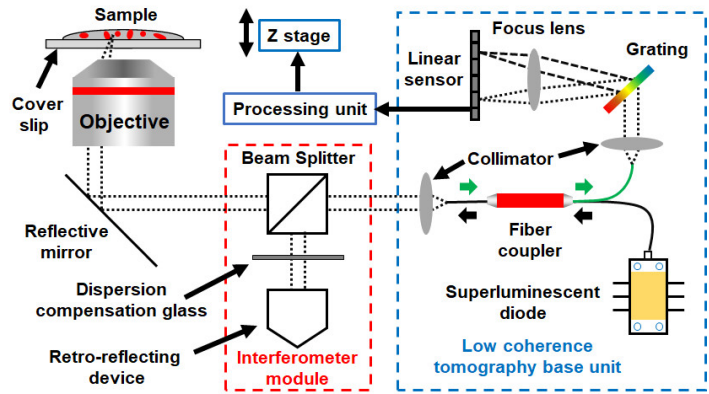

**Figure 5.** Low-coherence interferometry for reflective real-time autofocusing. A superluminescent diode is used as a low-coherence light source. The light illuminates the sample from a tilted incident angle. As such, most reflected light from the glass surface will not be coupled back to the interferometry system. The axial depth reflectivity profile (i.e., A-scan) is measured using a spectrometer. The recovered sample position is used to move the z stage or the objective lens. Adapted from Ref. [113].

One solution to this problem is to substantially reduce the light reflected from glass surfaces while keeping the sample scattering light relatively unchanged. Figure 5 demonstrates such a solution by using an off-axis configuration, where the light illuminates the sample at a tilted incident angle[113]. As such, the light directly reflected from the glass surface will not be coupled back to the interferometry system. In Figure 5, a broadband superluminescent diode is used as the low-coherence light source. The axial depth reflectivity profile (i.e., A-scan) is measured using a spectrometer in a Fourier-domain OCT setup. The sample position can be calculated by performing a Fourier transform of the captured spectrum and used to move the objective lens to the in-focus position.

Since OCT is sensitive to refraction index variations within the sample, this approach can handle transparent samples that may be challenging for the traditional focus map approach. The disadvantages, perhaps, are the complicated Fourier-domain OCT setup, the precise optical alignment, and the high maintenance of the system.

## 4. Real-time image-based autofocusing

The pre-scan focus map approach requires the acquisition of a z-stack for each focus point. The sample needs to be scanned to different x-y positions for acquiring multiple z-stacks to generate the focus map. In many WSI systems, the overhead time for generating the focus map is a substantial portion of the total scanning time. In this section, we will discuss several real-time image-based autofocusing approaches without the need for generating the focus map.

### 4.1. Independent dual sensor scanning

The traditional focus map approach uses the same image sensor to both survey the focus and acquire the image. In between two image acquisitions, there is a certain amount of 'dead time' to read out the data to the memory. As a result, the main camera cannot be used to survey the focus during this 'dead time'. An independent secondary image sensor has been proposed to survey the focus in parallel[6,114].

Figure 6 shows the principle and operation of this concept. In Figure 6(a), an independent camera, termed focusing sensor, is used to survey the focus while the main camera captures the high-resolution sample images. During the scanning process, the stage is in continuous motion and the motion blur is eliminated by using short pulses of light during imaging. As shown in Figure 6(b), the focusing sensor acquires three autofocus images, each at a slightly different focal plane. Based on these three images, the system calculates the optimal focus position and moves the sample to that focal plane[25], where the main camera takes a high-resolution image. When the main camera is reading out image data, the autofocusing is repeated for the next tile position to predict its optimal focal plane ahead. Since the stage is in continuous motion during this process, the captured three focus images only share a small region of overlap (Figure 6(c)). Only the overlapping region is used to calculate the correct focal position. The autofocusing performance of this system has been validated with various tissue sections[114]. The average focusing error is ~0.30 μm for the continuous motion scheme. Around 95% of tiles fall within the system's depth of field.

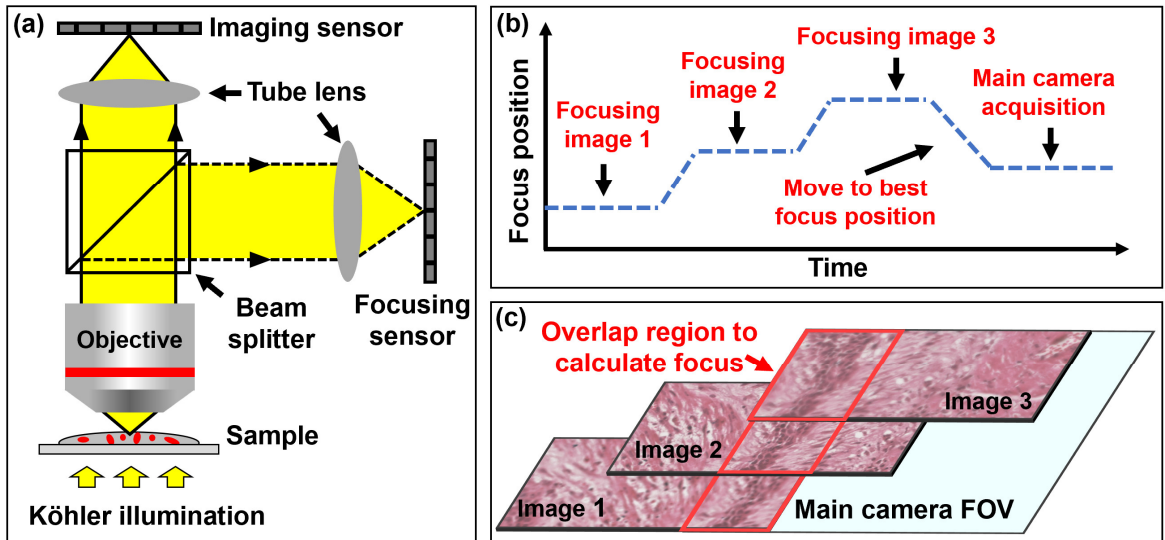

**Figure 6.** Independent dual sensor scanning for real-time image-based autofocusing. (a) The optical scheme, where a high-speed focusing camera is used to survey the focus in parallel with the main camera. (b) The focusing sensor acquires three autofocus images, each at a slightly different focal plane. The system calculates the optimal focus position and moves the sample to that focal plane, where the main camera takes a high-resolution image. (c) The stage is in continuous motion during this process and the captured three images only share a small region of overlap. Modified from Ref. [6].

## 4.2. Beam splitter array

In the independent dual sensor scanning scheme discussed above, multiple images are acquired to calculate the focus position when the sample is moved to different focal planes. In a patent published in 2010, Virag *et al.* proposed to use a beam splitter array to allow capturing images at different focal planes on the same image sensor[115]. Figure 7 shows the imaging setup, where the focusing optics comprises a main imaging camera and a secondary focusing camera. A beam splitter array is used to split and direct the light beam to different regions of the focusing sensor. As such, the system can capture images at multiple focal planes at the same time. The 45-degree semi-reflective surfaces in the beam splitter array are chosen to assure that all beams reflected by the surfaces have roughly the same intensities. With the image captured by the focusing sensor, a certain focus measure and fitting model can be used to infer the optimal focus position. In additional to autofocusing, this scheme can also be modified for real-time multiplane microscopy[116-119], which finds important applications in volumetric imaging of biological samples.

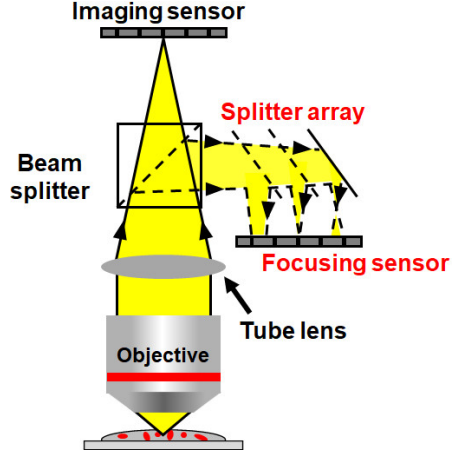

**Figure 7.** Beam splitter array for real-time image-based autofocusing. A beam splitter array is used to split and direct the light beam to different regions on the focusing sensor. As such, the system can capture images at multiple focal planes for determining the optimum focus position. Modified from Ref. [115].

### 4.3. Tilted sensor

The tilted sensor approach uses a tilted focusing sensor to image an oblique cross-section of the sample. The optimum focus position can be inferred by locating the peak of the contrast curve in real time. The concept of this approach was originally proposed in a patent by Dong *et al.* in 2005[120]. There are some further refinements and developments of this original concept by Philips[121-126] and Leica[127,128]. Arguably, it is one of the most successful autofocusing technologies employed in existing commercially available WSI systems.

Figure 8 shows the principle and operation of the tilted sensor concept. In Figure 8(a), the focusing sensor is tilted at $\theta$ angle with respect to the parfocal image plane. The imaging and focusing sensors can be either 2D area sensors or 1D linear sensors. The overlapping position between the focusing sensor and the parfocal imaging plane is termed 'parfocal point' in Figure 8(b). The focusing range is determined by $Z_{range}$. With a larger tilted angle, a longer focusing range can be expected.

During the scanning process, both sensors capture images of the sample. For each pixel of the captured data, a contrast value can be determined based on the surrounding pixel values. Consider a 1D image data $I(x)$ as an example, the contrast value $C(x)$ can be calculated via $C(x) = \sum_{m=-M}^{m=M} |I(x) - I(x-m)|$, where $m$ define the surrounding range for the calculation. A contrast curve can then be obtained by dividing the focusing sensor contrast value $C_{focus}$ by the imaging sensor contrast value $C_{image}$, as shown in Figure 8(c). The peak of the contrast curve determines the pixel having the highest contrast value, i.e., the best focal position. The parfocal point can also be plotted on the contrast curve. In Figure 8(c), the pixel distance $\Delta N$ between the parfocal point and the peak contrast point on the curve indicates a physical distance along the axial direction. This distance represents the distance between the current position of the objective lens and the optimal focus position of the objective lens, i.e., one needs to axially move the objective lens by this distance for best focusing. While the imaging sensor is centered at the field of view of the objective lens, the focusing sensor can be shifted away from the center of the field of view. As such, the focusing sensor 'sees' the image data before the imaging sensor 'sees' the same region.

Similarly, a 'volume camera' consisted of multiple linear CCDs coupled with fibers can be arranged with a tilted angle for autofocusing[129]. Bravo *et al* reported the use of 9 sensors coupled with fibers to acquire images at different focal planes for real-time image-based autofocusing[41].

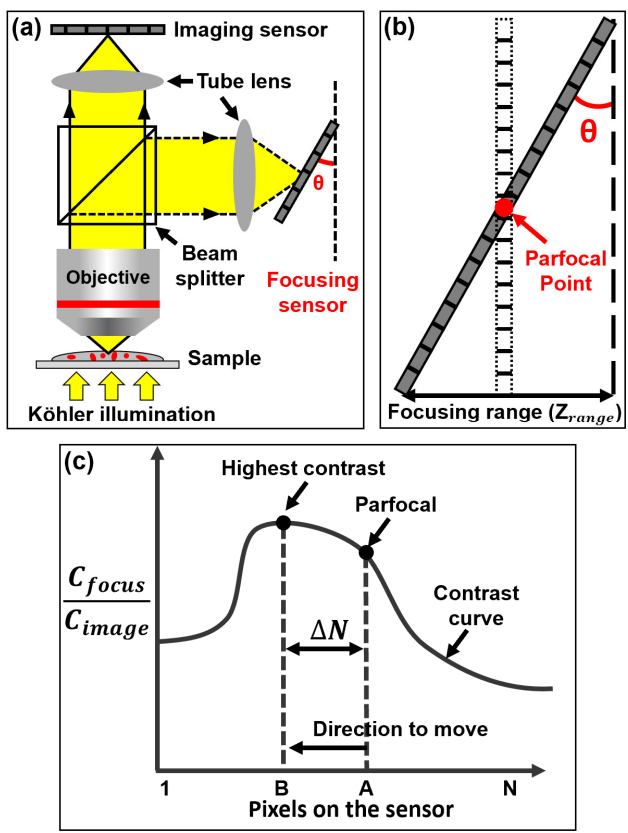

**Figure 8.** Tilted sensor for real-time image-based autofocusing. (a) The optical scheme, where a tilted sensor is used to infer the optimum focus position during the scanning process. (b) The overlapping position between the focusing sensor and the parfocal imaging plane is termed 'parfocal point'. (c) Contrast curve for determining the optimal focus position. The pixel distance ($\Delta N$) between the parfocal point and the peak contrast point indicates a physical distance by which one needs to adjust the objective lens for optimal focusing. Modified from Ref. [127].

### 4.4. Phase detection

Phase detection autofocusing has been used in most digital single-lens reflex cameras (DLSR)[130]. It is typically achieved by dividing the incoming light into pairs of images. It then measures the distance between the two images and infers the defocus amount. The 'phase' here is referred to the translational shift between the two images (or the phase shift in the Fourier domain).

Inspired by the phase detection concept in photography, we have developed an autofocusing add-on kit to perform WSI using a regular microscope[131]. As shown in Figure 9(a), two pinhole-modulated cameras are attached to the eyepiece for phase detection autofocusing. By adjusting the positions of the pinholes, one can effectively change the view angles through the two eyepiece ports. If the sample is placed at the in-focus position, the two captured images will be identical. If the sample is placed at an out-of-focus position, the sample will be projected at two different view angles, causing a translational shift in the two captured images. The translational shift is proportional to the defocus distance of the sample. Therefore, by identifying the translational shift of the two captured images via phase correlation, the optimal focal position of the sample can be recovered without a z-scan.

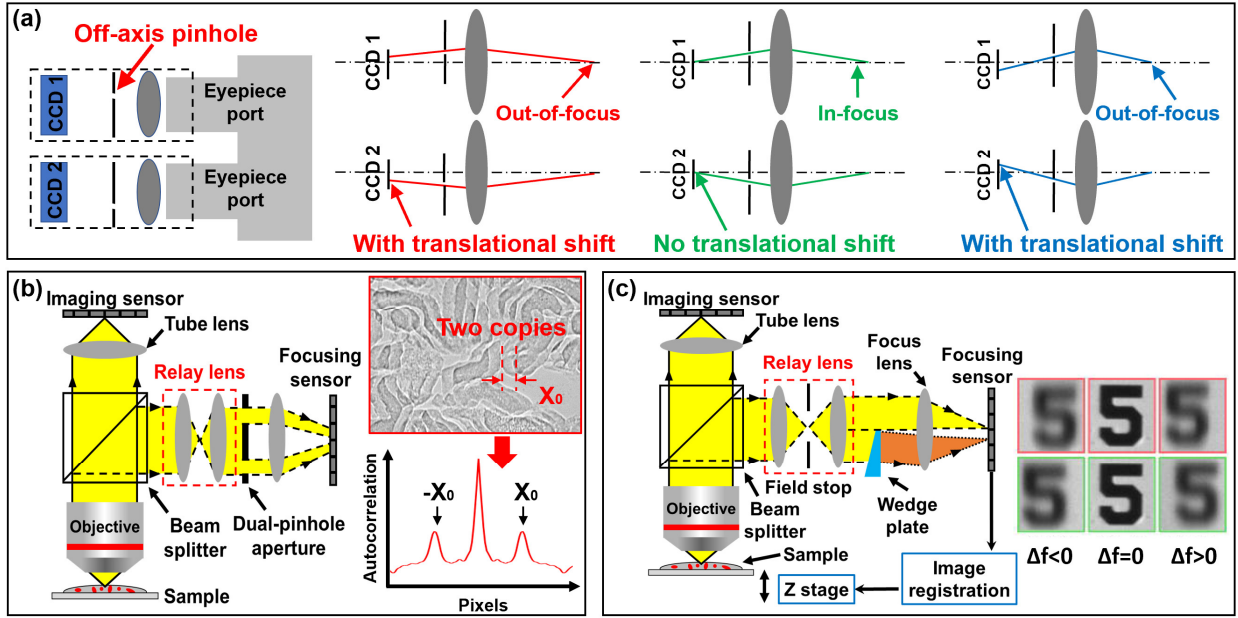

**Figure 9.** Phase detection for real-time image-based autofocusing. (a) Two pinhole-modulated cameras are attached to the eyepiece ports for phase detection autofocusing. If the sample is placed at an out-of-focus position, it will be imaged at two different view angles, causing a translational shift in the two captured images through the eyepiece ports. Modified from Ref. [131]. (b) A dual-pinhole mask is placed at the pupil plane for light modulation. The captured image from the focusing sensor contains two copies of the object and the defocus distance can be recovered based on the translational shift between the two copies. Modified from Ref. [132]. (c) A wedge plate is inserted into the pupil plane to direct half of the beam to a slightly tilted angle. As such, the captured image from the focusing sensor contains two copies of the sample separated by a certain distance. Similarly, the defocus distance can be recovered from the translational shift of the two copies. Modified from Ref. [133].

Figure 9(b) shows another autofocusing configuration using the phase detection concept[132]. A dual-pinhole mask is placed at the pupil plane to modulate the light from the sample. Instead of using two pinhole-modulated cameras, only one focusing sensor is used to capture the image modulated by the dual-pinhole mask. In this case, the captured image from the focusing sensor contains two copies of the sample and the translational shift of these two copies is proportional to the defocus distance. Inset of Figure 9(b) shows a raw image captured by the focusing sensor, where two copies of the sample can be seen from this image. The distance between the two copies can be recovered via autocorrelation analysis shown in Figure 9(b).

Figure 9(c) shows a similar phase detection scheme by Silvestri *et al.*[133]. Same as the dual-pinhole modulation approach, only one camera is used for the focusing purpose. A wedge plate is inserted into the pupil plane to direct half of the beam to a slightly tilted angle. As such, the captured image from the focusing sensor contains two copies of the sample separated by a certain distance. The defocus distance can be recovered from the translational shift of the two copies.

For the configurations shown in Figure 9(a) and 9(b), pinhole masks are used to restrict the light at the pupil plane. Therefore, they have relatively long autofocusing ranges. The system in Figure 9(c), on the other hand, has a short autofocusing range. Using the dual-pinhole mask does not prevent its applications in fluorescence microscopy. One can choose a beam splitter cube to direct the strong excitation light through the dual-pinhole mask. Weak fluorescence emissions from the sample can be directed to the imaging camera. The configuration in Figure 9(b) has been demonstrated for fluorescence WSI[132].

## 4.5. Dual-LED illumination

Dual-LED illumination has recently been demonstrated for single-frame autofocusing while the sample is in continuous motion[134-138]. Figure 10(a) shows one of the reported configurations where two near-infrared LEDs are placed at the back focal plane of the condenser lens for sample illumination[134]. These two LEDs illuminate the sample from two different incident angles and they can be treated as spatially coherent light sources. A hot mirror is used to direct the near-infrared light to the focusing sensor shown in Figure 10(a). As such, the captured image from the focusing sensor contains two copies of the sample separated by a certain distance. In particular, the focusing sensor is placed at a preset offset distance with respect to the imaging sensor. When the sample is placed at the in-focus position, the captured image from the focusing sensor contains two copies of the sample profile. Similar to the dual-pinhole mask approach, one can recover the defocus distance by identifying the separation of the two copies through autocorrelation analysis. The preset offset arrangement in Figure 10(a) is used to improve the accuracy of autocorrelation analysis when the defocus distance is small. It can also generate out-of-focus contrast for transparent specimens. If the sample motion direction is perpendicular to the direction of the translational shift, the autofocusing process can be implemented even with continuous sample motion. This dual-LED scheme has also been demonstrated for focus map surveying with only one main camera[136].

Figure 10(b1) shows a further development of the dual-LED approach using color multiplexed illumination[137,138]. In this scheme, a color LED array is used for sample illumination. For regular brightfield image acquisition, all LED elements are turned on as shown in the left part of Figure 10(b1). In between two brightfield acquisitions, a red and a green LED are turned on for color-multiplexed illumination. If the sample is placed at an out-of-focus position, the red and the green copy will be separated by a certain distance, as shown in the insets of Figure 10(b1). One can then identify the translational shift of the red- and green-channel images by maximizing the image mutual information or cross-correlation[139, 140]. The resulting translational shift is used for dynamic focus correction in the scanning process.

Figure 10(b2) shows an open-source WSI platform, termed OpenWSI, based on the color-multiplexed dual-LED autofocusing scheme[138]. This OpenWSI platform is built with low-cost, off-the-shelf components including a programmable LED array, a photographic lens, and a computer numerical control (CNC) router. Coarse axial adjustment is performed using the CNC router and the precise adjustment is performed using the ultrasonic motor ring within the photographic lens. The system has a resolution of ~0.7 μm using a 20X objective lens. It can acquire a whole slide image of 225 mm$^2$ region in ~2 mins. Since a programable LED array is used for sample illumination in this system, it can also be used for quantitative phasing imaging via Fourier ptychography.

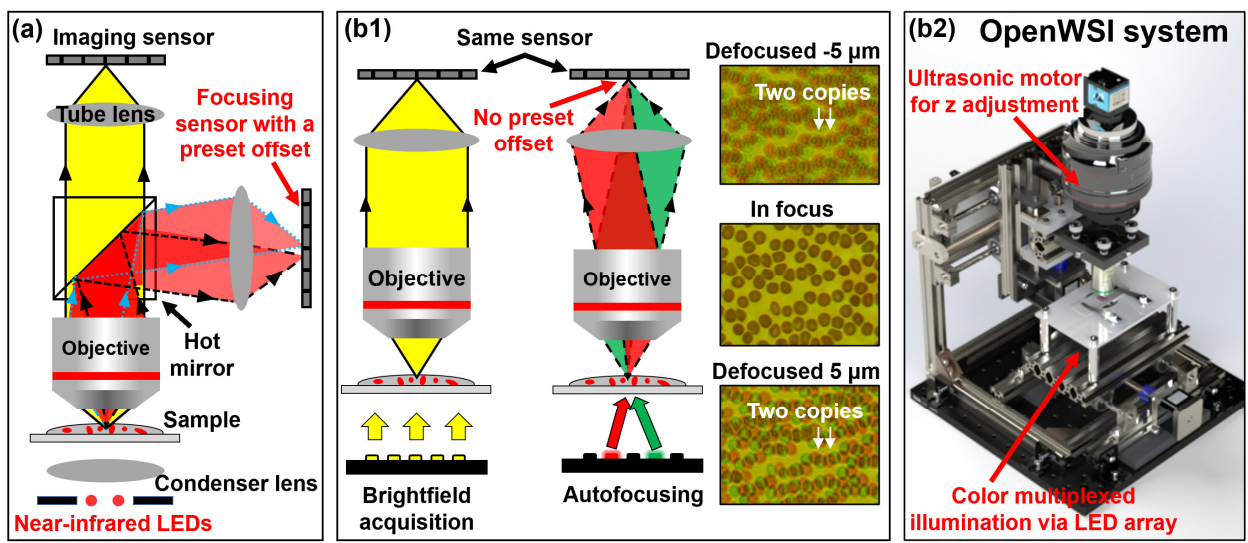

**Figure 10.** Dual-LED illumination for single-frame autofocusing. (a) Two near-infrared LEDs are placed at the back focal plane of the condenser lens for illuminating the sample from two different angles. A hot mirror is used to direct the near-infrared light to the focusing sensor with a preset offset. The defocus distance is related to the separation of the two-copy image captured by the focusing sensor. (b1) Color-multiplexed dual-LED illumination for single-frame autofocusing. A red and a green LED are turned on for generating a red and green copy on the color image sensor. (b2) OpenWSI system based on the color-multiplexed dual-LED autofocusing scheme. Modified from Ref.[138].

### 4.6. Deep learning approaches

Deep learning has been demonstrated as a powerful tool for solving inverse problems. With the advent of accelerated computing and deep learning frameworks such as TensorFlow and PyTorch, researchers have also explored various deep learning-based solutions for autofocusing[21,92,139-150]. As shown in Figure 11, the reported deep-learning solutions can be, in general, categorized into two groups.

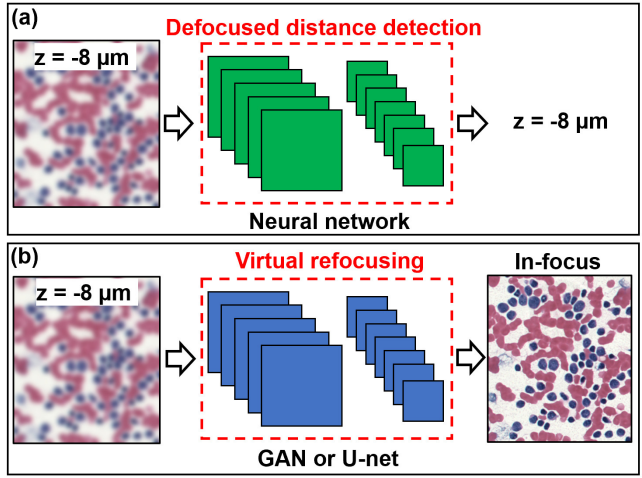

**Figure 11.** Deep learning approaches for autofocusing. (a) A neural network is trained to output the defocus distance from an input defocused image. (b) A neural network is trained to output an in-focus image based on the input defocused image.

The first group is to predict the defocus distance or to locate the out-of-focus regions based on one or more input defocused images[21,92-94,140,141,144,147,149,150]. For example, Jiang *et al*. employed a convolutional neural network (CNN) to estimate the defocus distance based on the transform- and multi-domain inputs[141]. By adding the Fourier spectrum and the autocorrelation of the spatial image as the input, the performance and the robustness can be improved compared to that only with the spatial image as the input. Dastidar *et al*. further improved the performance by using the difference of two defocused images as the input of the CNN[140]. Shajkofci *et al.* reported the use of a CNN-based sharpness function as the focus measure for three-shot autofocusing[147]. Pinkard *et al*. designed a fully connected Fourier neural network with the additional off-axis LEDs as the illumination source to predict the defocus distance[144]. Yang *et al*.[94] and Kohlberger *et al*.[21] have developed networks to quantify and localize the out-of-focus regions in WSI. The severity of the out-of-focus regions is treated as a classification problem with 30 classes[21].

The second group of developments is to output an in-focus image based on an input defocused image[142,143,146,148]. The network is, essentially, to perform blind deconvolution. Typically network architectures include U-net[151] and conditional generative adversarial network (cGAN)[152]. For example, Wu *et al*. have employed a cGAN to virtually refocus a two-dimensional fluorescence image onto user-defined three-dimensional (3D) surfaces by appending a pre-defined digital propagation matrix[148]. It has also been shown that a blurry microscopy image acquired at an arbitrary out-of-focus plane can be virtually refocused to the in-focus position[143].

## 5. Summary and discussion
High-content images are desired in many fields of biomedical research as well as in clinical applications. Accurate and high-speed autofocusing remains a challenge for WSI and automated microscopy. This work has reviewed and discussed various autofocusing techniques from existing patents and journal papers. The technical concepts, merits, and limitations of these methods are explained and discussed.

| Autofocusing approach | Advantages | Disadvantages |
|---|---|---|
| Focus map | • No or less intellectual property issue<br>• Require no additional optical hardware<br>• Can be used for different imaging modalities<br>• Robust and widely adopted for WSI | • Require a z-stack for each focus point<br>• Mechanical repeatability is critical for sample positioning<br>• Challenging to handle transparent specimens |
| Confocal pinhole | • High accuracy for locating the air-glass interface | • Require additional confocal optics<br>• Time-consuming for z-scan<br>• Reflection from other interfaces can be overwhelmed by the strong signal from the air-glass interface |
| Triangulation with oblique illumination | • High accuracy for locating the air-glass surface of a standard coverslip<br>• Real-time autofocusing | • Require additional illumination and detection optics<br>• Only work for living cells housed in imaging chambers with a standard coverslip. Cannot work for microscope slides or thick plastic dish. |
| Low-coherence interferometry | • Can handle transparent specimens<br>• Real-time autofocusing | • Expensive and complicated Fourier-domain OCT setup<br>• Precise optical alignment needed |
| Independent dual sensor scanning | • Real-time image-based autofocusing during continuous sample motion<br>• Effectively avoid the 'dead time' of camera readout | • Require a secondary area camera and pulsed illumination<br>• Require the acquisition of three images for autofocusing with a small overlapping portion<br>• Relatively short autofocusing range |
| Beam splitter array | • Real-time image-based autofocusing | • Require a secondary area camera<br>• Relatively short autofocusing range |
| Tilted sensor | • Real-time image-based autofocusing<br>• Fully compatible with linear and TDI image sensor<br>• Fast calculation via contrast curve<br>• One of the most successful techniques deployed in commercially available WSI systems | • Require a secondary focusing sensor<br>• A transparent sample may lead to a wrong autofocusing calculation since out-of-focus regions have a higher contrast |
| Phase detection | • Real-time image-based autofocusing<br>• Can handle transparent specimens via a preset offset of the focusing sensor | • Require additional camera(s) and relay optics for the pinhole mask<br>• Precise alignment needed for the pinhole mask<br>• Low-pass filtering of the pinhole mask may affect the accuracy of the correlation analysis |
| Dual-LED illumination | • Real-time image-based autofocusing<br>• Can be implemented with continuous sample motion<br>• Can handle transparent specimens<br>• Relatively long autofocusing range due to the use of partially coherent dual-LED illumination<br>• Cost effective and compatible with most automated microscope platforms | • Only work for regular 2D thin slides |
| Deep learning | • Allow single-shot autofocusing<br>• Require no additional optical hardware | • Relatively short virtual refocusing range<br>• Change of optical hardware may affect the autofocusing performance<br>• The system may fail for new features or new types of specimens that have not been trained before |

**Table 3.** Summary and comparison of different autofocusing techniques.

We summarize the advantages and disadvantages of these techniques in Table 3. Among these techniques, the focus map approach is the most adopted technique in existing WSI systems due to its simplicity and the absence of intellectual property issues. The tilted sensor approach is another very successful technique employed in current Leica and Philips WSI systems. The recent dual-LED approach provides a cost-effective solution to develop WSI systems that can be made broadly available and utilizable without loss of capacity. The deep learning approach, on the other hand, is an emerging direction for tackling autofocusing problems without hardware modification. Further work is desired for improving its robustness and the generalization capability of handling different types of specimens.

Some of the autofocusing techniques discussed here can also be employed in the augmented reality microscope system. For example, a secondary tilted sensor can be used for locating the optimal focus position in real-time. A motorized stage can be used to drive the main camera for capturing the in-focus, high-resolution sample images.

In the medical realm, one strategy taken by the National Cancer Moonshot initiative to fight cancer cooperatively is to create an image database for different cases and connect scientists and pathologists for online collaboration. Coupling an automated microscope system with a proper autofocusing technique has the potential to convert various biological specimens into high-content images and address the challenge of high-throughput microscopy.

**Acknowledgments**
Z. B. and C. G. contributed equally to this work. P. S. acknowledges the support of the Thermo Fisher Scientific Fellowship. K. H. acknowledges the support of NSF 1809047. G. Z. acknowledges the support of NSF 1700941 and NSF 2012140.